\documentclass[aps,pra,preprint,groupedaddress,showpacs]{revtex4}

\begin{document}

\title{Spin Bath Decoherence of Quantum Entanglement}

\author{Zafer Gedik}

\email{gedik@sabanciuniv.edu}

\affiliation{Faculty of Engineering and Natural Sciences, Sabanci
University, Tuzla, Istanbul 34956, Turkey}

\date{\today}

\begin{abstract}
We study an analytically solvable model for decoherence of a two
spin system embedded in a large spin environment. As a measure of
entanglement, we evaluate the concurrence for the Bell states
(Einstein-Podolsky-Rosen pairs). We find that while for two
separate spin baths all four Bell states lose their coherence with
the same time dependence, for a common spin bath, two of the
states decay faster than the others. We explain this difference by
the relative orientation of the individual spins in the pair. We
also examine how the Bell inequality is violated in the coherent
regime. Both for one bath and two bath cases, we find that while
two of the Bell states always obey the inequality, the other two
initially violates the inequality at early times.
\end{abstract}

\pacs{03.65.Yz, 75.10.Jm, 03.65.Ud, 03.67.Mn, 03.65.Ta, 03.65.Ud}

\maketitle

Entanglement, nonlocal quantum correlations between subsystems, is
not only one of the basic concepts in quantum
mechanics~\cite{schrodinger} but also central to quantum
computation and quantum information~\cite{nielsen}. Decoherence,
loss of phase relations between the states, is essential in
understanding how a quantum system becomes effectively
classical~\cite{zurek03}. Therefore, how an entangled system
undergoes decoherence or how the entanglement changes as a result
of interaction with the environment is an important issue and for
two entangled spins subject to quantum noise created by a bosonic
bath the problem has already been studied~\cite{yu, tolkunov}.

In this work, we concentrate on decoherence of two spins as a
result of an interaction with a spin bath. This problem is closely
related to electron spin dynamics, due to hyperfine interaction
with surrounding nuclear spins, in quantum dots~\cite{schliemann}.
Decoherence of various systems, including superconducting quantum
interference devices (SQUIDs) coupled to nuclear and paramagnetic
spins, can be described by similar models~\cite{prokofev}. Many
spin systems can exhibit interesting behaviors including parity
dependent decoherence where some nondiagonal elements of the
density matrix survive the initial decay of other entries due to
environmental spins~\cite{dobrovitski,melikidze}. For the central
spin model, which describes a localized spin coupled to a spin
bath, the quasiclassical equations of motion are
integrable~\cite{yuzbashyan}.

Quantification of entanglement is a major challenge in quantum
information theory. A well known measure for a pure state of a
pair of quantum systems is the von Neumann entropy or equivalently
the Shannon entropy of the squares of the Schmidt
coefficients~\cite{bennett}. The entropy of the partial density
matrix, which is obtained by tracing out one of the members from
the total density matrix, can be used to parametrize the
entanglement. For a pair of binary quantum objects (qubits) an
alternative parameter is the concurrence which is related to the
von Neumann entropy bijectively~\cite{hillwootters}. To quantify
the entanglement between the two spins, we are going to use the
concurrence because of its mathematical simplicity. Our main
results related to entanglement will turn out to be independent of
the choice of the measure.

Decoherence of the two spins can be viewed as a generation of
entanglement between the pair and the spin bath (or baths) and
hence any measure of the entanglement can also be used to
parametrize the decoherence. As the members of the pair lose their
entanglement with each other, they start to entangle with the bath
spins. What we are going to evaluate is the concurrence
corresponding to the entanglement between the two partners.

Our aim is to understand how two entangled spins lose their
correlation due to other spins interacting with them.  For this
purpose we start with a very simple model where we can observe
decoherence effects. The model Hamiltonian
\begin{equation}\label{ham}
H=c_{1z}\sum_{k=1}^{N_{1}}\hbar\omega_{1k}\sigma_{1kz}+
c_{2z}\sum_{k=1}^{N_{2}}\hbar\omega_{2k}\sigma_{2kz}
\end{equation}
describes two central spins, with $z-$component operators $c_{1z}$
and $c_{2z}$, coupled to bath spins represented by $\sigma_{nkz}$,
where $n=1,2$ labels the baths and $k=1,2,3,...,N_{n}$ labels the
individual spins. All spins are assumed to be 1/2 and $c_{1z}$,
$c_{2z}$, and $\sigma_{nkz}$ denote the corresponding Pauli
matrices. Hamiltonian (\ref{ham}) is a simple two spin
generalization of the model proposed by Zurek to study decoherence
in spin systems~\cite{zurek82}. First, we are going to consider
two different spin baths where each spin couples only one of them.
Later, we are going to examine how our results change when the
pair interacts with a single bath. Since the Hamiltonian
(\ref{ham}) involves only the $z-$components of the spins, it can
also be used to study decoherence of other two-state systems.

We are going to assume that at $t=0$, the central spins are not
entangled to the spin baths so that the state is in the product
form $|\Psi(0)\rangle=|\Psi_{c}(0)\rangle
|\Psi_{\sigma1}(0)\rangle|\Psi_{\sigma2}(0)\rangle $ where
\begin{equation}
|\Psi_{c}(0)\rangle=\left(a_{\uparrow\uparrow}|\uparrow\uparrow\rangle+
a_{\uparrow\downarrow}|\uparrow\downarrow\rangle+
a_{\downarrow\uparrow}|\downarrow\uparrow\rangle+
a_{\downarrow\downarrow}|\downarrow\downarrow\rangle\right)
\end{equation}
with obvious notation for the two spins and
\begin{equation}
|\Psi_{\sigma n}(0)\rangle=
\bigotimes_{k=1}^{N_{n}}\left(\alpha_{nk}|\uparrow _{nk}\rangle+
\beta_{nk}|\downarrow_{nk} \rangle\right)
\end{equation}
where $|\uparrow _{nk}\rangle$ and $|\downarrow_{nk} \rangle$ are
eigenstates of $\sigma_{nkz}$ with eigenvalues +1 and -1,
respectively, and $|\alpha_{nk}|^{2}+|\beta_{nk}|^{2}=1$. At later
times, the state is no more in the product form due to
entanglement of the pair with environmental spins but instead it
is given by
\begin{equation}\label{state}
\begin{array}{rl}
 |\Psi(t)\rangle=(a_{\uparrow\uparrow}&|\uparrow\uparrow\rangle
|\Psi_{\sigma1}(+t)\rangle|\Psi_{\sigma2}(+t)\rangle \\
+ a_{\uparrow\downarrow}&|\uparrow\downarrow\rangle
|\Psi_{\sigma1}(+t)\rangle|\Psi_{\sigma2}(-t)\rangle\\
+ a_{\downarrow\uparrow}&|\downarrow\uparrow\rangle
|\Psi_{\sigma1}(-t)\rangle|\Psi_{\sigma2}(+t)\rangle\\
+ a_{\downarrow\downarrow}&|\downarrow\downarrow\rangle
|\Psi_{\sigma1}(-t)\rangle|\Psi_{\sigma2}(-t)\rangle) \\
\end{array}
\end{equation}
where
\begin{equation}\label{bstate}
|\Psi_{\sigma
n}(t)\rangle=\bigotimes_{k=1}^{N_{n}}(\alpha_{nk}e^{-i\omega_{nk}t}|\uparrow
_{nk}\rangle+ \beta_{nk}e^{i\omega_{nk}t}|\downarrow_{nk}
\rangle).
\end{equation}
We are going to see that it is the randomness of the interaction
strengths and the expansion coefficients that will lead to
decoherence of the pair.

The total, central spin and the baths, density matrix is given by
$\rho(t)=|\Psi(t)\rangle\langle\Psi(t)|$ but what we are
interested in is the reduced density matrix which is obtained from
the former by tracing out the bath degrees of freedom as
$\rho_c(t)=Tr_{\sigma}\rho(t)$ where subscript $\sigma$ means that
trace is evaluated by summing over all possible $nk$ states. We
can write the resulting density matrix in the product basis
$\{|\uparrow\uparrow\rangle, |\uparrow\downarrow\rangle,
|\downarrow\uparrow\rangle, |\downarrow\downarrow\rangle\}$ as
\begin{equation}
\rho_{c}=\left(
\begin{array}{cccc}
  |a_{\uparrow\uparrow}|^{2} & a_{\uparrow\uparrow}a_{\uparrow\downarrow}^{*}r_{2} &
  a_{\uparrow\uparrow}a_{\downarrow\uparrow}^{*}r_{1}& a_{\uparrow\uparrow}a_{\downarrow\downarrow}^{*}r_{1}r_{2}\\
  a_{\uparrow\uparrow}^{*}a_{\uparrow\downarrow}r_{2}^{*} & |a_{\uparrow\downarrow}|^{2} &
  a_{\uparrow\downarrow}a_{\downarrow\uparrow}^{*}r_{1}r_{2}^{*} & a_{\uparrow\downarrow}a_{\downarrow\downarrow}^{*}r_{1} \\
  a_{\uparrow\uparrow}^{*}a_{\downarrow\uparrow}r_{1}^{*} & a_{\uparrow\downarrow}^{*}a_{\downarrow\uparrow}r_{1}^{*^{}}r_{2}&
  |a_{\downarrow\uparrow}|^{2}&a_{\downarrow\uparrow} a_{\downarrow\downarrow}^{*}r_{2}  \\
  a_{\uparrow\uparrow}^{*}a_{\downarrow\downarrow}r_{1}^{*}r_{2}^{*} & a_{\uparrow\downarrow}^{*}a_{\downarrow\downarrow}r_{1}^{*} &
  a_{\downarrow\uparrow}^{*}a_{\downarrow\downarrow}r_{2}^{*} & |a_{\downarrow\downarrow}|^{2} \\
\end{array}
\right)
\end{equation}
where $*$ means complex conjugation and decoherence factors
$r_{1}(t)$ and $r_{2}(t)$ are given by
\begin{equation}\label{rn}
r_{n}(t)=\prod_{k=1}^{N_{n}}\left(|\alpha_{nk}|^{2}e^{-i2\omega_{nk}t}+|\beta_{nk}|^{2}e^{i2\omega_{nk}t}\right).
\end{equation}
In general both expansion coefficients $\alpha_{nk}$, $\beta_{nk}$
and interaction strengths $\omega_{nk}$ are random. If the bath
spins point randomly at $t=0$ we can write the expansion
coefficients as $\alpha_{nk}=\cos(\theta_{nk}/2)e^{-i\phi_{nk}/2}$
and $\beta_{nk}=\sin(\theta_{nk}/2)e^{i\phi_{nk}/2}$, where
$\theta_{nk}$ and $\phi_{nk}$ are spherical polar coordinates
determining the direction of the spins, and we assume that the
angles $\theta_{nk}$ and  $\phi_{nk}$ have uniform distributions
in the intervals $[0,\pi]$ and $[0,2\pi]$, respectively. It is
possible to show that for sufficiently large $N_{n}$ values
$|r_{n}(t)|$ exhibits a Gaussian time dependence $e^{-a_{n}t^{2}}$
rather than exponential~\cite{zurek04}. In our case
\begin{equation}\label{an}
a_{n}=16\sum_{k}|\alpha_{nk}|^{2}|\beta_{nk}|^{2}|\omega_{nk}|^{2}.
\end{equation}
We are going to obtain several coherence factors given by
expressions similar to Eq.~(\ref{rn}). We first note that the
larger the interaction strengths $|\omega_{nk}|$, the faster the
decay. Secondly, for a given set of $\{\omega_{nk}\}$, the fastest
decoherence is attained when $|\alpha_{nk}|$ and $|\beta_{nk}|$
become equal.

To evaluate the concurrence~\cite{hillwootters}, we need to find
the time-reversed or spin-flipped density matrix $\rho_{c}$ which
is given by
\begin{equation}
\widetilde{\rho}_{c}=(\sigma_{y}\otimes\sigma_{y})\rho_{c}^{*}(\sigma_{y}\otimes\sigma_{y}).
\end{equation}
Here $\sigma_{y}$ is the Pauli spin matrix and $\otimes$ stands
for the Kronecker product, and $\rho_{c}^{*}$ is obtained from
$\rho_{c}$ via complex conjugation. We can write the spin-flipped
density matrix as
\begin{equation}
\widetilde{\rho}_{c}=\left(
\begin{array}{cccc}
  |a_{\downarrow\downarrow}|^{2} & -a_{\downarrow\uparrow}a_{\downarrow\downarrow}^{*}r_{2} &
  -a_{\uparrow\downarrow}a_{\downarrow\downarrow}^{*}r_{1}& a_{\uparrow\uparrow}a_{\downarrow\downarrow}^{*}r_{1}r_{2}\\
  -a_{\downarrow\uparrow}^{*}a_{\downarrow\downarrow}r_{2}^{*} & |a_{\downarrow\uparrow}|^{2} &
  a_{\uparrow\downarrow}a_{\downarrow\uparrow}^{*}r_{1}r_{2}^{*} & -a_{\uparrow\uparrow}a_{\downarrow\uparrow}^{*}r_{1} \\
  -a_{\uparrow\downarrow}^{*}a_{\downarrow\downarrow}r_{1}^{*} & a_{\uparrow\downarrow}^{*}a_{\downarrow\uparrow}r_{1}^{*}r_{2}&
  |a_{\uparrow\downarrow}|^{2}& -a_{\uparrow\uparrow}a_{\uparrow\downarrow}^{*}r_{2}  \\
  a_{\uparrow\uparrow}^{*}a_{\downarrow\downarrow}r_{1}^{*}r_{2}^{*} & -a_{\uparrow\uparrow}^{*}a_{\downarrow\uparrow}r_{1}^{*} &
  -a_{\uparrow\uparrow}^{*}a_{\uparrow\downarrow}r_{2}^{*} & |a_{\uparrow\uparrow}|^{2} \\
\end{array}
\right)
\end{equation}
The final step in evaluation of the concurrence $C$ is to find the
four eigenvalues $\{\lambda_{i}\}$ of the product matrix
$\rho\widetilde{\rho}_{c}$ in the decreasing order so that
\begin{equation}\label{concur}
C=\max\{0,\sqrt{\lambda_{1}}-\sqrt{\lambda_{2}}-\sqrt{\lambda_{3}}-\sqrt{\lambda_{4}}\}.
\end{equation}
We are going to evaluate the above expression for the Bell states
(Einstein-Rosen-Podolsky pairs)
\begin{equation}\label{bell}
\begin{array}{ll}
 |e_{1}\rangle& =\frac{|\uparrow\uparrow\rangle+|\downarrow\downarrow\rangle}{\sqrt{2}} \\
 |e_{2}\rangle& =\frac{|\uparrow\downarrow\rangle+|\downarrow\uparrow\rangle}{\sqrt{2}} \\
 |e_{3}\rangle& =\frac{|\uparrow\uparrow\rangle-|\downarrow\downarrow\rangle}{\sqrt{2}} \\
 |e_{4}\rangle& =\frac{|\uparrow\downarrow\rangle-|\downarrow\uparrow\rangle}{\sqrt{2}} .\\
\end{array}
\end{equation}
As we are going to see, the Bell states  have the property that
the concurrence is the same for all of them. In fact, any other
basis obtained from the Bell states by replacing the coefficients
$\pm1/\sqrt{2}$ with $e^{i\theta}/\sqrt{2}$ ($\theta$ being a real
number) has the same property.

For two baths, the concurrence, which is the same for all of the
Bell states, turns out to be
\begin{equation}\label{con2}
C=|r_{1}||r_{2}|.
\end{equation}
Since, $r_{1}(0)=r_{2}(0)=1$, the concurrence is also unity at
$t=0$. On the other hand, both $r_{1}$ and $r_{2}$, and hence the
concurrence, decay with time and vanish. For the special case
where only one of the spins, say the first one, interacts with a
spin bath so that $r_{2}(t)=r_{2}(0)=1$, we still observe a decay
in the concurrence. This is an expected result because the
entangled pair must be treated as a single system rather than
individual spins.

We next consider the case where both spins undergo decoherence due
to interaction with the same spin bath so that the Hamiltonian
becomes
\begin{equation}
H=\hbar \sum_{k=1}^{N}(\omega_{1k}c_{1z}+
\omega_{2k}c_{2z})\sigma_{kz}
\end{equation}
This time the state at $t=0$ is given by
$|\Psi(0)\rangle=|\Psi_{c}(0)\rangle |\Psi_{\sigma}(0)\rangle$
where
\begin{equation}
|\Psi_{\sigma}(0)\rangle= \bigotimes_{k=1}^{N}(\alpha_{k}|\uparrow
_{k}\rangle+ \beta_{k}|\downarrow_{k} \rangle).
\end{equation}
Similar to the two bath case, we can write the density matrix as
\begin{equation}
\rho_{c}=\left(%
\begin{array}{cccc}
  |a_{\uparrow\uparrow}|^{2} & a_{\uparrow\uparrow}a_{\uparrow\downarrow}^{*}r_{2} &
  a_{\uparrow\uparrow}a_{\downarrow\uparrow}^{*}r_{1}& a_{\uparrow\uparrow}a_{\downarrow\downarrow}^{*}r_{12}^{+}\\
  a_{\uparrow\uparrow}^{*}a_{\uparrow\downarrow}r_{2}^{*} & |a_{\uparrow\downarrow}|^{2} &
  a_{\uparrow\downarrow}a_{\downarrow\uparrow}^{*}r_{12}^{-} & a_{\uparrow\downarrow}a_{\downarrow\downarrow}^{*}r_{1} \\
  a_{\uparrow\uparrow}^{*}a_{\downarrow\uparrow}r_{1}^{*} & a_{\uparrow\downarrow}^{*}a_{\downarrow\uparrow}r_{12}^{-*}&
  |a_{\downarrow\uparrow}|^{2}& a_{\downarrow\uparrow}a_{\downarrow\downarrow}^{*}r_{2}  \\
  a_{\uparrow\uparrow}^{*}a_{\downarrow\downarrow}r_{12}^{+*} & a_{\uparrow\downarrow}^{*}a_{\downarrow\downarrow}r_{1}^{*} &
  a_{\downarrow\uparrow}^{*}a_{\downarrow\downarrow}r_{2}^{*} & |a_{\downarrow\downarrow}|^{2} \\
\end{array}%
\right)
\end{equation}
where decoherence factors are given by
\begin{equation}
r_{n}(t)=\prod_{k=1}^{N}\left(|\alpha_{k}|^{2}e^{-i2\omega_{nk}t}+|\beta_{k}|^{2}e^{i2\omega_{nk}t}\right) \\
\end{equation}
and
\begin{equation}
r_{12}^{\pm}(t)=
\prod_{k=1}^{N}\left(|\alpha_{k}|^{2}e^{-i2(\omega_{1k}\pm\omega_{2k})t}+
 |\beta_{k}|^{2}e^{i2(\omega_{1k}\pm\omega_{2k})t}\right). \\
\end{equation}
As we have discussed in the paragraph after Eq.~(\ref{an}), the
larger the interaction strengths, the faster the decay. Therefore,
we should compare $\omega_{1k}+\omega_{2k}$  with
$\omega_{1k}-\omega_{2k}$. If all of the interaction constants
$\omega_{nk}$ have the same sign, $r_{12}^{+}(t)$ goes to zero
faster than $r_{12}^{-}(t)$. For the special case,
$\omega_{1k}=\omega_{2k}$ for all $k$, $r_{12}^{-}(t)$ does not
decay at all but remains constant.

After finding the spin-flipped density matrix
$\widetilde{\rho}_{c}$ and eigenvalues of the product
$\rho\widetilde{\rho}_{c}$, we can evaluate the concurrence for
each of the Bell states. In single bath case the Bell states
exhibit different decay rates with the concurrence expressions
\begin{equation}\label{con1}
\begin{array}{rl}
  C_{1}=C_{3} & =|r_{12}^{+}|\\
  C_{2}=C_{4} & =|r_{12}^{-}|.\\
\end{array}
\end{equation}
Although we have obtained this two by two grouping of the Bell
states in terms of the concurrence, any other measure, like the
von Neumann entropy, which depends upon the eigenvalues of the
density matrices will yield the same result. Equations
(\ref{con2}) and (\ref{con1}) show that the concurrence, which is
a measure of entanglement is given by nothing but the coherence
factor.

We can explain the different results for one bath and two baths
decoherence processes in terms of different characters of the Bell
states. When the spins interact with separate baths, relative
orientation of spins is not important because the only difference
between the up and down configurations is complex conjugation of
the coherence factor and it is the modulus of the coherence factor
which enters the concurrence expression. On the other hand, in the
single bath case there is no simple relation between the opposite
spin terms. In $|e_{1}\rangle$ and $|e_{3}\rangle$ states, spins
are always parallel while in $|e_{2}\rangle$ and $|e_{4}\rangle$
states, they are always antiparallel. That is why two groups have
different decoherence behaviors.

Finally, we are going to examine how the Bell inequality is
violated in the quantum regime and how it is satisfied in the
classical domain~\cite{bell}. The Bell inequality, or in fact
inequalities are satisfied if there exists a local realistic
theory~\cite{einstein}. There are a large number of Bell
inequalities, all resulting from local realistic assumptions, but
following Ref.~\cite{clauser} we will focus our attention on the
quantity
\begin{equation}\label{bellin}
S=E(\theta_{1},\theta_{2})-E(\theta_{1},\theta_{2}')+
E(\theta_{1}',\theta_{2}')+E(\theta_{1}',\theta_{2}),
\end{equation}
where the correlation function $E(\theta_{1},\theta_{2})$ is given
by,
\begin{equation}
E(\theta_{1},\theta_{2})=Tr\{\hat{c}_{1}(\theta_{1})
\otimes\hat{c}_{2}(\theta_{2})\;\rho_{c}\},
\end{equation}
with
\begin{equation}
\hat{c}_{i}(\theta_{i})=c_{iz}\cos\theta_{i}+c_{ix}\sin\theta_{i}.
\end{equation}
The Bell inequality is violated if $|S|>2$. In calculating whether
the inequality is violated, the choice of the angles $\theta_{1}$
and $\theta_{2}$ is crucial. It is known that not all entangled
states violate a Bell inequality~\cite{werner,popescu}. That is
why $\theta_{i}$s must be chosen carefully. In our case, for the
Bell states $\{|e_{i}\rangle\}$, we can find the corresponding
$\{S_{i}\}$ expressions. To simplify the equations, for a given
set of angles $\theta_{1}$, $\theta_{2}$, $\theta_{1}'$, and
$\theta_{2}'$ we will introduce the notation
\begin{equation}
\begin{array}{rl}
  A=&(\cos\theta_{1}\cos\theta_{2}-\cos\theta_{1}\cos\theta_{2}'\\
  &+\cos\theta_{1}'\cos\theta_{2}'+\cos\theta_{1}'\cos\theta_{2})\\
&\\
B=&(\sin\theta_{1}\sin\theta_{2}-\sin\theta_{1}\sin\theta_{2}'\\
&+\sin\theta_{1}'\sin\theta_{2}'+\sin\theta_{1}'\sin\theta_{2}) \\
\end{array}
\end{equation}
so that, for two separate baths,
\begin{equation}\label{bells}
\begin{array}{rr}
  S_{1}= & A+B\; \Re\{r_{1}r_{2}\}\\
  S_{2}=& -A+B\; \Re\{r_{1}r_{2}^{*}\}\\
  S_{3}=& A-B\; \Re\{r_{1}r_{2}\}\\
  S_{4}=& -A-B\; \Re\{r_{1}r_{2}^{*}\}\\
\end{array}
\end{equation}
where $r_{1}$ and $r_{2}$ are again given by Eqn.~(\ref{rn}), and
$\Re\{z\}$ denotes the real part of the complex number $z$. For a
single bath very similar expressions hold. In this case
$\Re\{r_{1}r_{2}\}$ and $\Re\{r_{1}r_{2}^{*}\}$ are replaced by
$\Re\{r_{12}^{+}\}$ and $\Re\{r_{12}^{-}\}$, respectively.

The angles $\theta_{1}$, $\theta_{2}$, $\theta_{1}'$, and
$\theta_{2}'$ can take arbitrary values. We are going to pick up a
particular set for which $\{S_{i}\}$ are easy to calculate. We
will assume that $\theta_{1}=0$, $\theta_{2}=\pi/4$,
$\theta_{1}'=\pi/2$, and $\theta_{2}'=3\pi/4$. For this choice of
angles, $A=B=\sqrt{2}$. At $t=0$ where all decoherence factors are
unity, for both two bath and single bath cases, $S_{2}$ and
$S_{3}$ vanish, and therefore they satisfy the Bell inequality
$|S|\leq2$. When the system becomes completely incoherent so that
all coherence factors vanish, again for both two bath and single
bath cases, we obtain $|S_{2}|=|S_{3}|=\sqrt{2}$. Although there
is an increase in $|S|$ values, the inequality is still satisfied.
In $|e_{1}\rangle$ and $|e_{4}\rangle$ states however, the Bell
inequality is violated at $t=0$, since
$|S_{1}|=|S_{4}|=2\sqrt{2}$. As the decoherence factors vanish,
they both decay to $\sqrt{2}$. For two baths, $|S_{1}|$ and
$|S_{4}|$ exhibit the same decay rate. In single bath case, the
corresponding factors coming from decoherence for $|e_{1}\rangle$
and $|e_{4}\rangle$ states are given by $\Re\{r_{12}^{+}\}$ and
$\Re\{r_{12}^{-}\}$, respectively. As we have discussed above, the
two factors decay at different rates.

In conclusion, using the concurrence and the Bell inequality, we
demonstrated that a pair of entangled spins show different
decoherence behaviors when the spins interact with a common spin
bath or separate baths.  Some entangled states can be more
vulnerable than others. For example, two entangled electrons in
the same quantum dot will have a different coherence
characteristics than two in separate dots. Recent proposal by
Beenakker \emph{et al.} for the creation of entangled
electron-hole pairs  might be an interesting system to look for
such decoherence effects~\cite{beenakker}.

The author would like to thank Cihan Sa\c{c}l{\i}o\u{g}lu for
fruitful and valuable discussions. This work has been partially
supported by the Turkish Academy of Sciences, in the framework of
the Young Scientist Award Program (MZG/
T{\"U}BA-GEB{\.I}P/2001-2-9).

\end{document}